\documentclass[journal]{IEEEtran}

\usepackage{multirow}
\usepackage{graphicx}
\usepackage{tabularx}
\usepackage{booktabs}
\usepackage{xurl}

\newenvironment{myitemize}[1][]{
	\begin{list}{$\bullet$}
		{
			\setlength{\leftmargin}{5mm}
			\setlength{\parsep}{1mm}
			\setlength{\topsep}{0mm}
			\setlength{\itemsep}{0mm}
			\setlength{\labelsep}{1.5mm}
			\setlength{\itemindent}{0mm}
			\setlength{\listparindent}{5mm}
	}}
	{\end{list}}

\ifCLASSINFOpdf
\else
\fi
\begin{document}
%
\title{From Plate to Production: Artificial Intelligence in Modern Consumer-Driven Food Systems}
%
%
%


\author{Weiqing~Min, \textit{Senior Member IEEE},
        Pengfei~Zhou,
        Leyi~Xu,
        Tao~Liu,
        Tianhao~Li,
        Mingyu~Huang,
        Ying~Jin,
        Yifan~Yi,
        Min~Wen,
        Shuqiang~Jiang, \textit{Senior Member IEEE},
        Ramesh~Jain, \textit{Fellow IEEE}
        \thanks{Weiqing Min, Pengfei Zhou, Leyi Xu, Tao Liu, Tianhao Li, Mingyu Huang, Ying Jin, Yifan Yi, Min Wen and Shuqiang Jiang are with the Key Laboratory of Intelligent Information Processing, Institute of Computing Technology, Chinese Academy of Sciences, Beijing, 100190, China, and also with the University of Chinese Academy of Sciences, Beijing 100049,
China (e-mail:sqjiang@ict.ac.cn).}
        \thanks{Ramesh Jain is with University of California, Irvine, CA, 92697, USA (e-mail:rcjain@uci.edu).}
        }

%
%

\markboth{Journal of \LaTeX\ Class Files,~Vol.~0, No.~0, month~year}%
{Shell \MakeLowercase{\textit{Min et al.}}: From Plate to Production: Artificial Intelligence in Modern Consumer-Driven Food Systems}

\maketitle

\begin{abstract}
Global food systems confront the urgent challenge of supplying sustainable, nutritious diets in the face of escalating demands. The advent of Artificial Intelligence (AI) is bringing in a personal choice revolution, wherein AI-driven individual decisions transform food systems from dinner tables, to the farms, and back to our plates. In this context, AI algorithms refine personal dietary choices, subsequently shaping agricultural outputs, and promoting an optimized feedback loop from consumption to cultivation.  Initially, we delve into AI tools and techniques spanning the food supply chain, and subsequently assess how AI subfields—encompassing machine learning, computer vision, and speech recognition—are harnessed within the AI-enabled Food System (AIFS) framework, which increasingly leverages Internet of Things, multimodal sensors and real-time data exchange. We spotlight the AIFS framework, emphasizing its fusion of AI with technologies such as digitalization, big data analytics, biotechnology, and IoT extensively used in modern food systems in every component. This paradigm shifts the conventional “farm to fork” narrative to a cyclical “consumer-driven farm to fork” model for better achieving sustainable, nutritious diets. This paper explores AI's promise and the intrinsic challenges it poses within the food domain. By championing stringent AI governance, uniform data architectures, and cross-disciplinary partnerships, we argue that AI, when synergized with consumer-centric strategies, holds the potential to steer food systems toward a sustainable trajectory. We furnish a comprehensive survey for the state-of-the-art in diverse facets of food systems, subsequently pinpointing gaps and advocating for the judicious and efficacious deployment of emergent AI methodologies.

\end{abstract}


%
\IEEEpeerreviewmaketitle

\section{Introduction}
Historically, the emergence of agriculture marked one of humanity's most transformative societal revolutions. In a world before agriculture, societies were largely dictated by the whims of nature, consuming what was immediately available. However, as agriculture took root and flourished, it catalyzed a sequence of advancements—from the birth of systematic food transportation and intricate processing techniques to sophisticated preparation and delivery mechanisms. This agricultural metamorphosis didn't just alter our diets. It further reshaped societal structures, resulting in settled communities and giving birth to civilizations.  This was all a result of advances in technology that continued to feed transformations.

Yet, the challenges and responsibilities borne by our modern food systems extend far beyond the agricultural fields. Today, food stands at the nexus of our health and happiness, environment, and intricate societal frameworks. It shapes societies in many implicit and explicit ways.  It bears significant weight in achieving the United Nations Sustainable Development Goals (SDGs) and realizing the Paris Agreement's ambitions. Remarkably, each of the 17 SDGs finds its relevance, either directly or indirectly, in our global food systems~\cite{hassoun2022fourth}. For us to stride confidently towards these benchmarks, we need to reimagine and accordingly redesign food systems that not only promote but champion sustainable healthy diets—ones that holistically nurture human well-being while ensuring minimal environmental impact, and remain universally available, accessible, affordable, and desirable~\cite{FFS-Report2020}.  Food systems should also promote equity and harmony in societies as environs in those 17 SDGs.

Our current food systems, however, are not even inching closer to this vision. From grappling with public health crises, like the triple burden of malnutrition, to confronting rampant food insecurity, the issues are manifold~\cite{micha20202020}. Further compounding these challenges are the intertwined threats of climate change, environmental degradation, and unchecked social exploitation~\cite{FFS-Report2020}. Events such as the COVID-19 pandemic, regional conflicts, and economic downturns have also highlighted the fragility and vulnerabilities embedded within our food supply chains~\cite{laborde2020covid, timmer2022food}.

In confronting challenges, technological innovations always provide hope. History serves as a testament to this. The mechanization revolution, for instance, heralded a new age of agriculture. Electrification saw the dawn of refrigeration, which extended food shelf-life and dramatically reduced wastage. The advent of telecommunications birthed precision agriculture, allowing farmers to monitor and manage their fields remotely and with unprecedented accuracy.

As we enter the fourth industrial revolution, Artificial Intelligence (AI) emerges with resounding potential. Techniques like deep learning have already made significant strides~\cite{lecun2015deep}. Considering its applications in detecting diseases in crops through image recognition, advanced neural networks, when coupled with drones, offer real-time monitoring of agricultural fields, ensuring early detection of pests or anomalies~\cite{rs15092450, YU2021119493}. Fundamental AI models, akin to those revolutionizing protein folding in biology~\cite{bommasani2021opportunities, jumper2021highly}, are being deployed for optimizing crop genetics, ensuring we cultivate strains that are both nutritious and resilient to changing environmental conditions~\cite{wang2020deep}. Furthermore, AI-driven robotic systems promise automation in tedious tasks, from weed removal to fruit picking, ensuring precision and reducing human drudgery~\cite{bonaccorsi2017highchest}. In the realm of food safety, machine learning algorithms analyze vast datasets, predicting potential outbreaks and ensuring timely interventions~\cite{vilne2019machine,harrison2014using}. Personalized nutrition, a concept once relegated to the annals of science fiction, is now within reach, with AI curating diets tailored to individual genetic profiles and health needs, reminiscent of its transformative impacts in personalized medicine~\cite{us2020, denny2021precision}.

From enhancing crop yields and ensuring food quality to streamlining resource use, bolstering food safety, and crafting personalized food choices, AI's promises span the entire spectrum of food systems~\cite{mcclements2021building, MinJLRJ19}. When integrated with other cutting-edge technologies like the Internet of Things (IoT), robotics, and blockchain, its potential multiplies manifold. Yet, for AI to deliver on its full promise, a synergistic environment encompassing robust policies, state-of-the-art infrastructure, and societal behavioral shifts is indispensable~\cite{herrero2020innovation}.

\begin{figure}[htbp]
\centering
\includegraphics[trim=2.5cm 0 2.5cm 0, width=0.3\textwidth]{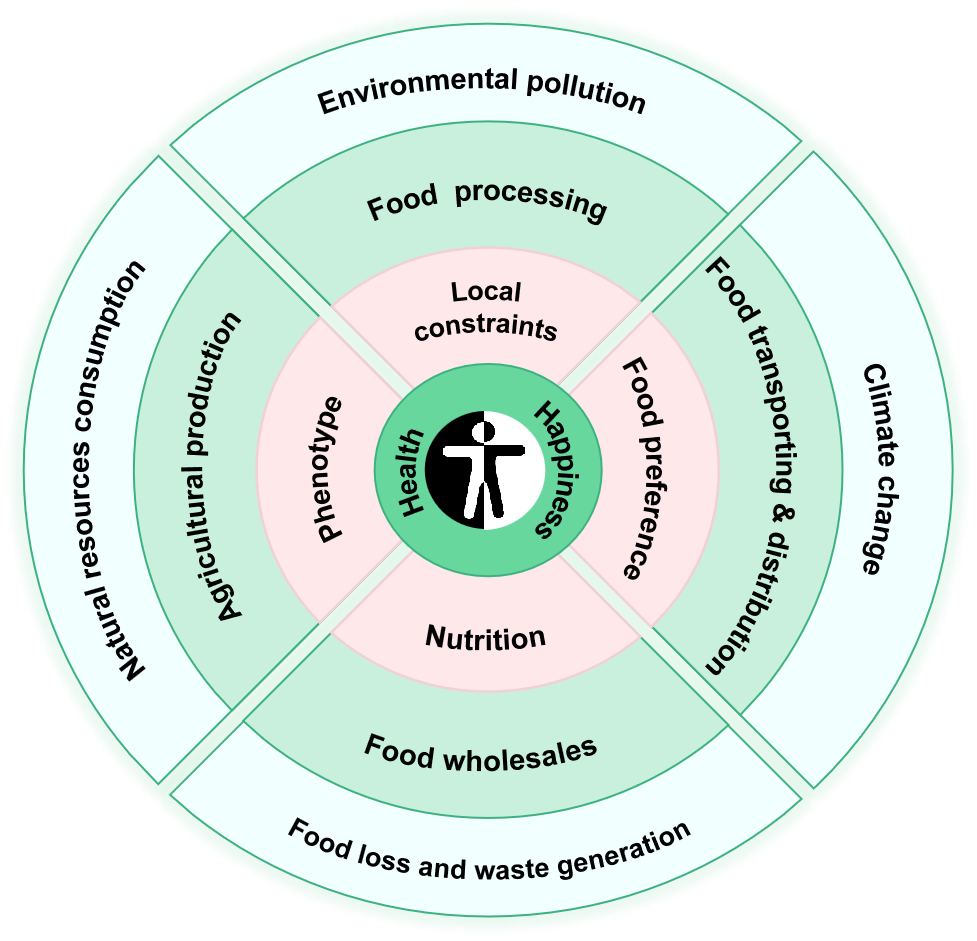}
\caption{The impact of individual choices on the environment.}
\label{impact}
\end{figure}

At the heart of these systems is the consumer, whose choices and feedback loops play a pivotal role in shaping the food landscape. As shown in Fig.~\ref{impact}, the shifts in consumption patterns, driven by factors ranging from health considerations to environmental imperatives, send important messages throughout the system, echoing the consumer-centric transformation that agriculture underwent in its early days. What happened at the beginning of agriculture is likely to happen at the speed of internet now and bring a major transformation in food systems again.  Through AI and associated technologies, we have the tools to make all sectors of food systems not just efficient but also responsive to dynamic consumer needs.

Embarking on this exploration, this paper introduces a framework of AI-enabled Food Systems (AIFS). We then journey through AI's potential in reshaping the availability, accessibility, affordability, and desirability of food in every stage of food systems, from consumption to production, processing, transportation
and storage of food. Finally, we spotlight the roadblocks and future research avenues. This narrative stands distinct from other reviews~\cite{friedlander2020artificial, bao2022artificial, MARVIN2022344}, offering a holistic perspective on AIFS from a consumer-driven perspective. But before we delve deep, a brief primer on AI sets the foundation.

\section{Overview of AI}
\begin{figure*}[htbp]
\centering
\includegraphics[trim=2.5cm 0 2.5cm 0, width=1\textwidth]{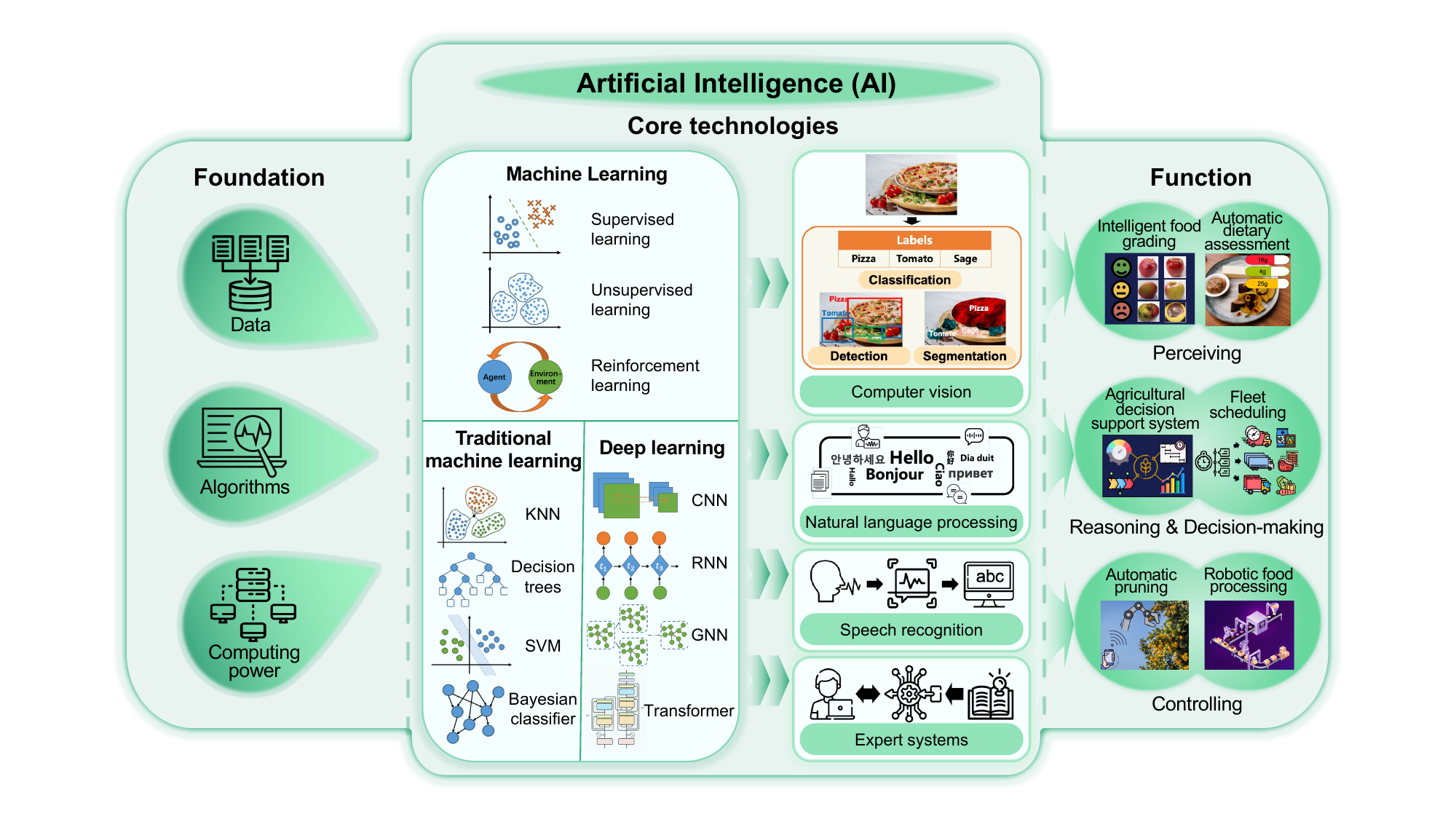}
\caption{Overview of AI. AI technologies contain computer vision, natural language processing, speech recognition, expert systems and others. It is built upon three key pillars, namely data, algorithms and computing power, and can serve functions that would normally require human intelligence like perceiving, reasoning, decision-making and controlling.}
\label{Overview_of_AI}
\end{figure*}

Artificial Intelligence (AI) is a subfield of computer science, aiming at building intelligent machines that can compute how to act as humans do~\cite{russell2010artificial}.  
Fig.~\ref{Overview_of_AI} illustrates the overview of AI. With the development of Machine Learning (ML), especially Deep Learning (DL), AI enters a third wave and achieves great improvement in computer vision, natural language processing, speech recognition, expert systems and other domains. It is built upon three key pillars, namely data, algorithms and computing power, and can serve functions that would normally require human intelligence like perceiving, reasoning, decision-making and controlling.

\subsection{Foundation of AI}
Data, algorithms and computing power represent three foundational pillars of AI. AI aims to learn potential rules, associations and patterns from massive data. Therefore, rapid growth in the amount of data could promote the modeling capability of AI. As the core of AI, algorithms are constantly developing based on large-scale data and open-source AI frameworks. Computing power from hardware has taken on a supporting role in the evolution of AI. The concurrent progress in cloud computing and edge computing is bringing new sources of computing power besides traditional infrastructures such as high-performance GPUs and specialized AI accelerators.

\subsection{AI Core Techniques}

\textbf{Machine Learning (ML)}  is devoted to building learning methods for machines, where models are learned based on sample data (e.g., training data) to make predictions or decisions~\cite{jordan2015machine}. Traditional ML methods need more human intervention such as data preprocessing, such as   K-Nearest Neighbors (KNN), and Support Vector Machines (SVM). In contrast, DL is currently becoming a hotspot among ML techniques~\cite{lecun2015deep}. It is based on multilayer neural networks connected by neurons, extending classical ML by adding more ``depth'' into models and transforming the data using various functions that allow data representation in a hierarchical way.
Different specialized neural network architectures are developed to capture specific properties of data, such as spatial locality caught by Convolutional Neural Networks (CNN), sequence information caught by Recurrent Neural Networks (RNN) and their variants, e.g., Long Short-Term Memory (LSTM), information exchange caught by Graph Neural Networks (GNN)  and context dependence caught by Transformers. 

Based on the learning types, ML can be categorized into supervised learning, unsupervised learning and reinforcement learning. In supervised learning, ML models are trained on examples with corresponding ground-truth labels to enable the machine to learn rules and predict the correct labels for new examples. As one branch, incremental learning and continual learning train models on a stream of annotated examples to continuously learn rules on new scenarios without forgetting previous rules, making models more generalized and suitable for real-world applications~\cite{de2021continual}. Unsupervised learning means training models without annotations, where models learn representations from unlabeled data and discover inherent data structure, such as clustering and autoencoders~\cite{hinton1999unsupervised}. Recently, self-supervised learning has become a hot research topic that contributes to the real-world application of unsupervised learning techniques~\cite{liu2021self}. Reinforcement Learning (RL)  forces an agent to interact with the ``environment'' strategically by receiving rewards continuously that reflect how well it is doing~\cite{li2017deep}. The interaction between agents and the ``environment'' is similar to that between humans and the environment, and RL can thus be considered as a set of learning frameworks for Artificial General Intelligence (AGI).  

\textbf{Computer Vision (CV) } focuses on deriving and processing information from images or videos, and teaching computers to understand visual inputs like human's eyes. Image classification, object detection and image segmentation are three typical CV tasks~\cite{russell2010artificial}. Image classification refers to predicting the class of an object represented by one given image. The main idea is to extract features from the images via feature extractors and classify the images based on the extracted features. Based on image classification, object detection further provides the position of objects via regression methods, which means simultaneously locating and classifying all the objects in an image. In contrast, image segmentation partitions an image into multiple regions by assigning the semantic label to each pixel of the image. The middle panel of Fig.~\ref{Overview_of_AI} shows one example of three CV techniques for automatic dietary assessment~\cite{Wang-ADA-TIFST2022}.

\textbf{Natural Language Processing (NLP)} focuses on theories and methods for effective communication between humans and computers with natural language. It offers computers the ability to read, understand and derive meaning from human languages. Common NLP tasks include text classification, named entity recognition, automatic summarization and dialogue robots. Pre-training models like BERT proved the effectiveness of pretrained Transformers on various NLP tasks~\cite{devlin2018bert}. Recently, Large Language Models (LLMs) like ChatGPT and GPT-4 based on the decoder of a similar transformer architecture further pushed forward the limits of understanding and generation~\cite{zhao2023survey}. With text collection and analysis, NLP methods benefit applications in food systems like food product design~\cite{brooks2022use} and food waste reduction~\cite{mishraUseTwitterData2018}.

\textbf{Speech Recognition}  is the ability of a machine or program to identify words that people speak and convert them into readable text. The speech recognition system consists of two stages: encoding and decoding. The audio signal is taken as the input of the model. The system encodes speech signals by extracting features and then uses a search algorithm to calculate their scores to select the best word sequence as the recognition result. For example, speech recognition can be used for improving cooking experiences as the smart kitchen assistant~\cite{angara2017foodie} and detecting the food quality~\cite{IYMEN2020102527}.

\textbf{Expert Systems} consist of a knowledge base and an inference engine. It uses knowledge representation and knowledge reasoning methods to simulate complex problems that can be solved by human experts in the domain, assisting users in making decisions. Different expert systems from various domains in the food systems have been developed for applications like agricultural decision~\cite{zhai2020decision} and food processing~\cite{sun2019recent}.

\subsection{Functions of AI}
The wide applications of AI in food systems originate from its capacity to allow machines to possess perceiving, reasoning, decision-making and control that only humans naturally have.  (1) \textbf{Perceiving}. The perceiving capability of AI is to understand the surrounding environment by interpreting information provided by sensors with different modalities, such as vision, hearing and touch~\cite{russell2010artificial}. The perception of things could promote the utilization of surrounding information. For example, AI technology can intervene in the process of crop growth~\cite{valente2019fast} and defective food identification~\cite{tbk2022detect} via image analysis. (2) \textbf{Reasoning \& Decision-making}. AI could infer new conclusions from known knowledge, and thus assist decision-making in the food systems. For example, different decision support systems have been widely employed to cover various agricultural applications including water resource management and climate change adaptation~\cite{zhai2020decision} for precision agriculture. (3) \textbf{Controlling}. AI can take over more than physical tasks by instructing robots to get involved in food production and processing. For example, intelligent robots could assist automatic food packaging for convenient meal storing based on  AI algorithms~\cite{ummadisingu2022cluttered}.

\section{AI-enabled food systems}
\begin{figure*}[htbp]
\centering
\includegraphics[trim=2.5cm 0 2.5cm 0,width=0.8\textwidth]{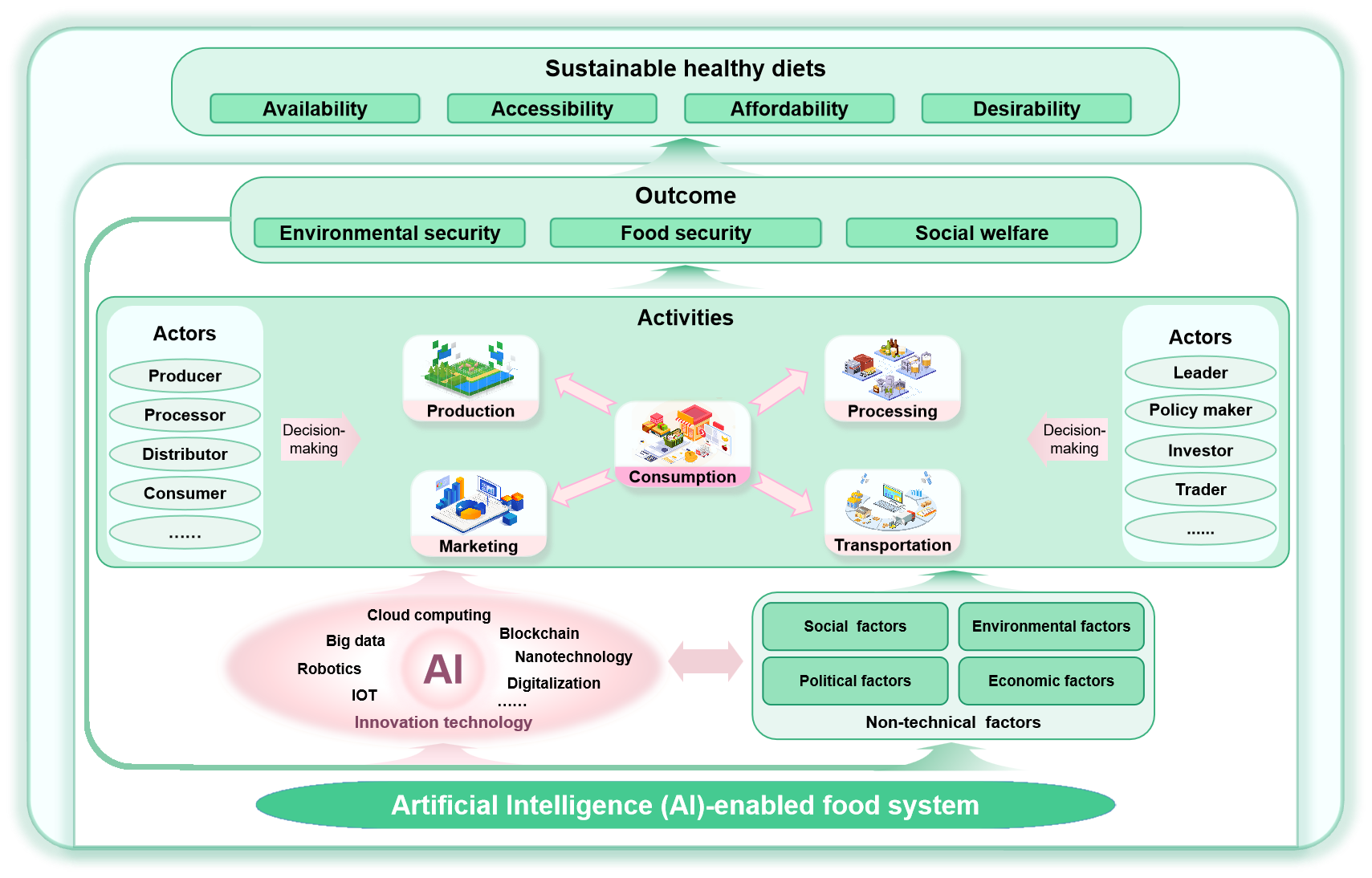}
\caption{AI-enabled food systems. The combination of AI and consumer-driven food systems gives rise to Al-enabled Food Systems (AIFS): under the right enabling political, social, economic, and natural environment, AIFS utilizes AI and other technologies (e.g., big data, IoT, robotics and biotechnology) to empower the entire range of actors and activities in the consumer-driven food systems from table to farm to improve the availability, accessibility, affordability and desirability of produced food for consumers toward sustainable healthy diets. In consumer-driven food systems, the requirements from food consumption drive the transition of other stages of food systems from food production to marketing, where AI can play positive roles in their transition.}
\label{AIFS_FS}
\end{figure*}

As shown in Fig.~\ref{AIFS_FS}, consumer-driven food systems encompass the entire range of actors and their interlinked activities, the broader social, economic and natural environments where these activities are embedded, along with their social, economic and environmental outcomes~\cite{nguyen2018sustainable}. These activities relate to growing, harvesting, processing, packaging, transporting,  marketing, and consuming of food originating from agriculture, forestry, fisheries and diverse food industries. In consumer-driven food systems, consuming is centric, and is the major driver to change other stages to satisfy the consumption need. For example, different consumption patterns drive what kinds
of food and products are produced. In order to make healthy diets affordable to all in the consumption, we should reduce the cost of producing, processing and delivering nutrient-rich foods. The actors consist of farmers, producers, processors, distributors, consumers, policymakers, investors, etc. A transition of food systems should ultimately involve all the activities, actors and their interactions. The goal of transformed food systems should ensure the availability, accessibility, affordability and desirability of the food for everyone seeking sustainable healthy diets.

The marriage of AI and food systems gives rise to AI-enabled Food Systems (AIFS): under the right enabling political, social, economic, and natural environment, AIFS utilizes AI and other technologies (e.g., big data, IoT, robotics, cloud computing, blockchain and nanotechnology) to empower the entire range of actors and activities in the consumer-driven food systems from table to farm to accelerate the positive transition of food systems toward sustainable healthy diets. From a macro perspective, food systems exhibit complexity, nonlinearity, dynamism, and unpredictability. AI is capable of understanding their interconnections and providing the capacity of decision support at the system level to make the food systems more autonomous and intelligent~\cite{klerkx2020dealing}. From a micro perspective, AI technologies coupled with other technologies, such as big data, IoT, blockchain and nanotechnology can optimize every stage of food systems to improve the availability, accessibility, affordability and desirability of produced food for consumers in the right enabling environment (e.g., basic physical infrastructures and digital systems, policy reformation and incentives, effective governance, unified regulatory and ethical standards).

Fig. \ref{AIFS_FS} shows the framework of AIFS. There are more interactions between technical innovation and non-technical political, social, economic and environmental factors. For example, the natural environment provides the resources needed for technological development and is in turn affected by AI technology. AI has positive impacts on the food systems under the premise that these interactions should be positive. It integrates and enhances both prevalent and emerging technologies in food systems, such as big data, IoT, robotics, nanotechnology, and biotechnology, enabling every stage of the AIFS from farm to fork.

The outcomes of AIFS contribute to food security, environmental security and social welfare to support sustainable food systems, and can also provide feedback to make AI be adapted to meet social, economic and environmental sustainability toward sustainable healthy diets. For example, in order to guarantee the environmental sustainability of food systems, AI models should be environmentally friendly by reducing the high energy cost and carbon footprint generated from massive computational resource requirements. Note that AIFS goes beyond a purely AI technical view, especially in the stage of consumption with complex social, political, cultural and individual factors. AI must be accompanied by these factors that enable their deployment in AIFS~\cite{herrero2021articulating}.

\section{AIFS for sustainable healthy diets}
In order to make AI-enabled food systems achieve a sustainable healthy diet, we next introduce how AI can benefit each stage of food systems from production to disposal to improve the availability, accessibility, affordability and desirability of the food to everyone in detail. Considering the central role of consuming in consumer-driven food systems, we first introduce AI-enabled food consuming.

\begin{table*}[htbp]
\newcommand{\tabincell}[2]{\begin{tabular}{@{}#1@{}}#2\end{tabular}}
\footnotesize
  \centering
  \caption{Summary of AI-enabled representative tasks in AIFS}\label{tab1}
  \scalebox{1}{
    \begin{tabular}{rllcl}
    \toprule
    \multicolumn{1}{c}{Activities} & Representative tasks & AI methods & \multicolumn{1}{c}{Purpose} & \multicolumn{1}{l}{Representative works} \\
    \midrule
    \multicolumn{1}{c}{\multirow{6}{*}{Consuming}} & Food recommendation & ML & \multicolumn{1}{c}{\multirow{6}{*}{\tabincell{c}{Affordability and \\ desirability of food}}}  & ref.~\cite{min2019food,chen2021personalized} \\
        & Visual shopping assistant & CV, ML &  & ref.~\cite{klasson2020using} \\
          & Smart kitchen assistant & Speech Recognition  &  & ref.~\cite{angara2017foodie} \\
          & Personalized nutrition & ML &  & ref.~\cite{verma2018challenges,zeevi2015personalized} \\
          & Dietary assessment & CV, ML &  & ref.~\cite{Wang-ADA-TIFST2022,myers2015im2calories, lu2020artificial} \\
          & Food safety monitoring & ML (ANN)  &  & ref.~\cite{yang2021machine} \\
    \midrule
    \multicolumn{1}{c}{\multirow{9}{*}{Production}}  & Plant breeding & ML & \multicolumn{1}{c}{\multirow{9}{*}{Availability of food}} & ref.~\cite{harfouche2019accelerating} \\& Crop yield prediction & DL (CNN, LSTM) & & ref.~\cite{Jiaxuan-DGP-AAAI2017} \\
          & \multicolumn{1}{l}{Pest and disease  diagnostics} & CV, DL &  & \multicolumn{1}{l}{ref.~\cite{Xinda-PDR-TIP2021}} \\
          & Fruit harvesting & DL & & \multicolumn{1}{l}{ref.~\cite{liu2018robust}} \\
          & Farming environment management & ML&     & ref.~\cite{LEALFILHO2022121662,doorn2021artificial} \\
          & Animal behavior and health monitoring & CV, ML&  & ref.~\cite{matthews2017automated} \\
          & Weed intervention and management & DL (CNN)&  & \multicolumn{1}{l}{ref.~\cite{sa2017weednet}} \\
      
          & Protein sequences design and function prediction & DL (GAN, GCN) & & \multicolumn{1}{l}{ref.~\cite{repecka2021expanding, alipanahi2015predicting}}\\
          & New biological structure design & DL (CNN, RNN) & & \multicolumn{1}{l}{ref.~\cite{eastman2018solving}}\\ 
    \midrule
    \multicolumn{1}{c}{\multirow{11}{*}{\tabincell{c}{Processing \\ \&Packaging}}} & Meat quality assessment & ML  & \multicolumn{1}{c}{\multirow{11}{*}{\tabincell{c}{Accessibility and \\ desirability of food}}}  & \multicolumn{1}{l}{ref.~\cite{penning2020machine}} \\
          & Dust explosion prevention & ML (ANN)&  & \multicolumn{1}{l}{ref.~\cite{arshad2021modelling}} \\
          & Fruit grading and sorting & CV, ML&  & ref.~\cite{wang2021advanced,dhiman2022fruit} \\
          & Food drying process controlling & ML, Expert System &    & \multicolumn{1}{l}{ref.~\cite{sun2019recent}} \\
          & Seafood compound detection & ML (SVM)&  & \multicolumn{1}{l}{ref.~\cite{tan2019quantitative}} \\
          & Sugar sample analysis & ML (SVR)&  & \multicolumn{1}{l}{ref.~\cite{da2022handheld}} \\
          & Heating process optimization & ML&     & ref.~\cite{yang2021mechanistic,ogundele2022machine} \\
          & Package air-tightness diagnostics & DL (CNN)&  & \multicolumn{1}{l}{ref.~\cite{polyakov2020industrial}} \\
          & Food package recycling & ML&  & \multicolumn{1}{l}{ref.~\cite{kroell2022optical}} \\
          & Food quality monitoring and sorting & ML, DL&  & \multicolumn{1}{l}{ref.~\cite{van2008kinetic,hassoun2023food}} \\
          & Robotic poultry processing & DL&  & \multicolumn{1}{l}{ref.~\cite{ahlin2022robotic, mitsioni2021modelling}} \\
    \midrule
    \multicolumn{1}{c}{\multirow{8}{*}{\tabincell{c}{Transporting \\ \&Packaging}}} & Supplier selection & ML  (Bayesian model, ANN) & \multicolumn{1}{c}{\multirow{8}{*}{Accessibility of food}} & ref.~\cite{zhangModelCombiningBayesian2019,stephenDecisionSupportModel2016} \\
          & Distribution planning & ML&  & ref.~\cite{tingMiningLogisticsData2014,tufanoMachineLearningApproach2022} \\
          & Fleet scheduling & ML  (Genetic Algorithm)&  & ref.~\cite{wangMultiobjectiveOptimizationDelivering2018} \\
          & Supply chain cooperation & DL  (LSTM), RL&  & ref.~\cite{maoCreditEvaluationSystem2018,tangCooperationMultistageGame2005} \\
          & Food product tracing & ML&  & ref.~\cite{alfianImprovingEfficiencyRFIDbased2020,wongTechnicalSustainabilityCloudBased2021d} \\
          & Demand prediction & \multicolumn{1}{l}{ML  (Markov model), NLP} & & ref.~\cite{kappelmanOptimalControlDynamic2021,mishraUseTwitterData2018} \\
          & Food quality monitoring & \multicolumn{1}{l}{ML  (SVM, SVR, KNN)}&  & ref.~\cite{ gorzelanyModellingMechanicalProperties2022,choDeterminationHassAvocado2020} \\
          & Deterioration detection & ML  (SVM, KNN, ANN)&  & ref.~\cite{radyComparisonDifferentOptical2020, soniHyperspectralImagingMachine2022} \\
    \midrule
    \multicolumn{1}{c}{\multirow{4}{*}{Marketing}}  &Food product interation &DL (CNN)& \multicolumn{1}{c}{\multirow{4}{*}{ }} & ref.~\cite{philp2022predicting}\\
          &  Product attributes determination & ML&  & ref.~\cite{bi2022consumer} \\
          & Dynamic pricing & ML, RL &  & ref.~\cite{burman2021deep} \\
          & Sales forecasting & ML &  & ref.~\cite{tsoumakas2019survey} \\    
    \midrule
    \multicolumn{1}{c}{\multirow{6}{*}{Disposing}} & Defect detection \& grading & DL (CNN) & \multicolumn{1}{c}{\multirow{6}{*}{Accessibility of food}} & \multicolumn{1}{l}{ref.~\cite{tbk2022detect, tadhg2002inspect}} \\
          & By-product selection & ML (PCA) &  & \multicolumn{1}{l}{ref.~\cite{sab2021by}} \\
          & Cold-chain appliance management & DL (CNN) &  & \multicolumn{1}{l}{ref.~\cite{bonaccorsi2017highchest}} \\
          & Cafeteria demand prediction & ML (ANN) &  & \multicolumn{1}{l}{ref.~\cite{Faezi2021dining}} \\
          & Food fresh state detection &  DL (CNN) & & ref.~\cite{Lie2019household,Sanghi2022fresh} \\
          & Kitchen waste management & DL (CNN, RNN) & & \multicolumn{1}{l}{ref.~\cite{Zhang2021MSW,seo2021prediction}} \\

    \bottomrule
    \end{tabular}}%
  \label{tab:addlabel}%
\end{table*}%


\subsection{Food Consuming}
In the past few decades, sustainability of food consumption has become an emerging issue for the economy, environment and public health, which mainly involves the affordability and desirability of food~\cite{peeters2018obesity}. AIFS solves these issues and enables various applications for individual food choice, such as nutrition assessment, food recommendation and food safety monitoring~\cite{oliveira2021applicability}. AI technologies are progressively reshaping how consumers purchase, prepare, and maintain their diets~\cite{seo2020sensory}, where food safety runs through the whole process. AI-driven solutions empower consumers by refining their purchasing decisions for health and sustainability, facilitating innovative approaches to food preparation, and offering personalized dietary assessments and recommendations for optimal health outcomes. The following sections elucidate the role of AI in these areas, highlighting its impact on everyday choices, from grocery shopping to meal preparation and dietary management.

\subsubsection{AI for Food Purchasing and Preparation}
The choices of consumers may have certain blindness in food purchasing, deviating from the optimal food choice for sustainability and health. AI-enabled food recommendations can help provide consumers with food candidates to offer support for decision-making to meet the needs of sustainability and personalized health~\cite{min2019food,chen2021personalized}. In addition, consumers can also benefit from CV-based assistive technologies when they are shopping. For example, a multi-view generative model can accurately recognize grocery items for visually impaired consumers via retrieving images in a database~\cite{klasson2020using}. Food preparation such as recipe creation is supported by the computational creativity of AI~\cite{marin2019recipe1m}. In addition to identifying ingredients and reconstructing recipes based on visual features of dishes, multi-attribute theme models can fulfill various applications including flavor analysis and recipe recommendation for cooking procedures~\cite{min2017delicious}. For newly vulnerable populations like refugees, they can use ingredient substitution to find ingredients that match their recipes to prevent themselves from feeling disconnected due to cultural and nutritional differences~\cite{camarena2020artificial}.

\subsubsection{AI for Diet}
As the central role of consumer-driven food systems, diet directly affects human health. Since choosing food is a multi-faceted and personal-dependent process, food recommendation is extremely necessary~\cite{elsweiler2022food}. To determine the health effect of diets, nutritionists use controlled feeding studies, which are prohibitively expensive and contain substantial biases and errors~\cite{subramanian2014persistent}. AI solutions deployed on wearable devices have been proposed for managing food intake and monitoring energy expenditure~\cite{7801947}. Combined with various nutrition and health data, AI-based methods can automatically analyze consumers’ diets to provide precise nutrition for enhancing personalized health and wellness, resulting in precision nutrition~\cite{verma2018challenges}.  The goal of precision nutrition is to match individual features (e.g., genotypic and phenotypic features, food preference and nutrition) with the right foods (e.g., food composition and structures) to achieve desired physiological health~\cite{GAN2019675}. As one example, ML models can accurately predict personalized postprandial blood glucose responses by integrating various factors, such as food records, gut microbiome and anthropometrics~\cite{zeevi2015personalized}. Based on DL models and multimodal data from various wearable devices, automatic dietary assessment can be developed to rapidly and precisely assess the nutrition of food~\cite{myers2015im2calories,thames2021nutrition5k,shao2023vision}, and can further make personalized healthy decisions for hospitalized patients via capturing what people are eating and calculating the nutritional content of foods~\cite{lu2020artificial}. It is also feasible for diabetics to use a dietary assessment system to track their daily meals and obtain medical advice for maintaining their health~\cite{zhou2020cmrdf}. There have been web applications for food logging and analysis, which could provide users beneficial information about their dietary choices~\cite{10.1145/1630995.1631001}. Of note, dietary assessment requires high performance of food recognition models, and more advanced DL models have been recently developed to improve the recognition performance~\cite{wang2022ingredient,jiang2020multi}. DL can achieve ingredient recognition and food categorization simultaneously with reliable performance, or predicting ingredients and cooking methods together in food recognition, which could assess dietary nutrition for recipe retrieval and ~\cite{luo2023ingredient, chen2016deep, min2016being}. In addition, DL can also be employed for the identification of fruits and vegetables, enhancing the efficiency and accuracy of tasks in food processing~\cite{ min2023vision}.

Food safety in food consumption is vital to guarantee human health. AI has been used to track and detect food quality to deal with various food safety issues such as food poisoning and foodborne viruses. For example, ML methods enable the non-destructive detection of food pathogens via precisely recognizing the color changes on paper substrate~\cite{yang2021machine}. Furthermore, AI-based big data analysis platforms have been widely used to track food quality and safety throughout the complex food supply chain to avoid the outbreak of foodborne diseases~\cite{vilne2019machine}. As an example, AI is used to track and identify the sources of contaminated food by the New York City Health Department~\cite{harrison2014using}.

In consumer-driven food systems, the requirements from food consumption drive the transition of other stages of food systems from food production to marketing, where AI can play a positive role in their transition.

\subsection{Food Production}
For sustainable healthy diets, the main goal of food production is to make more nutrient-rich foods available for consumers. Therefore, we mainly introduce how AI can enhance food production to improve the availability of food. Current agricultural production is facing a wide range of challenges, such as low crop yields, declining soil health, and low use efficiency of chemical fertilizer, loss of food due to plant and animal disease~\cite{shahzad2019crop}. Riding on the development of sensors and IoT, sensory data collected by various sources, such as unmanned aircraft systems and satellites can provide a massive amount of data with various types, known as big data~\cite{wolfert2017big}. AI technologies coupled with big data can help build the agricultural decision support system, agricultural expert system and agricultural predictive analytic system to fine-tune complex factors to improve the overall efficiency of food production, belonging to the realm of precision agriculture~\cite{liu2020industry}. Given production mainly contains the activities associated with plants and animals, we first discuss AI-enabled methods for plant-oriented and animal-oriented food production. Another emerging revolutionary way of food production is food synthetic biology for biomanufacturing, which can convert renewable raw materials into important food components and functional food in the factory with much less land and less pollution, such as alternative proteins. Therefore, we also discuss recent works in AI-assisted food synthetic biology.

\subsubsection{AI for Plant-oriented Food Producation}
Breeding crops for higher yield and better adaptability to different climates is vital to ensure sustainable food production. AI can accelerate plant breeding via multi-omics big data integration combined with ensemble learning~\cite{salman2023crop}. In crop production systems, AI can help develop more accurate models to guide inputs such as water and fertilizers, assess crop status, and automatically plan effective and timely interventions in response to the change of crop conditions, improve soil quality and health indices~\cite{ANDRADE2021195}. Benefitting from the integration of CV and ML technologies, AI utilizes different types of imagery (e.g., RGB-D, multi-spectral and hyperspectral) to enable non-destructive prediction of plant growth and crop field~\cite{Jiaxuan-DGP-AAAI2017}, crop seeds screening~\cite{valente2019fast}, crop health monitoring~\cite{DadsetanPWHH21}, pest and disease diagnostics~\cite{ferentinos2018deep}, weed intervention and management~\cite{sa2017weednet}. Besides crop management, AI has also been widely applied in fruit and vegetable production such as fruit and vegetable counting for automatic harvesting~\cite{liu2018robust}, disease recognition and diagnosis~\cite{Xinda-PDR-TIP2021}, ripeness determination and quality assessment~\cite{taheri2020smart}. For example, plant disease diagnosis is very critical for increasing crop production. Most methods heavily rely on either the molecular assay or observations by plant protectors, which are complicated, time-consuming, or prone to errors. Combined with large-scale collected plant disease datasets, DL methods can automatically and accurately detect plant disease in real-time~\cite{Xinda-PDR-TIP2021}. To enable efficient plant production, environmental management is also important. ML-enabled analysis methods are capable of detecting climate change~\cite{LEALFILHO2022121662}, identifying agricultural soil properties~\cite{liakos2018machine}, optimizing the water sustainability~\cite{doorn2021artificial}, predicting planting date and harvest date~\cite{Robotics-Duckett-UK2018}. Furthermore, agricultural robots with advanced DL models can help improve the efficiency of crop harvesting and achieve automatic picking~\cite{stella2023can, Zhuoling-DAS-HRI2020}.

\subsubsection{AI for Animal-oriented Food Producation} It is important to achieve optimal and sustainable animal farming, which is not easy due to complex conditions of health, safety and behavior. AI can be used for improving animal health and welfare via environment monitoring and control~\cite{gautam2021machine}, animal behavior monitoring~\cite{matthews2017automated}, and animal identification and sorting~\cite{hu2020cow}. In aquaculture, AI can help realize automatic life information acquisition, aquatic product growth regulation and decision-making, fish disease diagnosis, aquaculture environment perception and regulations, and sustainable exploitation of natural fishery resources~\cite{gladju2022applications}, resulting in precision aquaculture~\cite{o2019precision}. Taking animal disease detection as one example, the poultry industry has consistently faced the menace of Newcastle Disease (ND). ND not only severely impacts the health of poultry, but also leads to changes in the acoustic features of their vocalizations. Leveraging this acoustic change, researchers introduced the deep poultry vocalization network for early ND detection~\cite{cuan2022automatic}. The integration of audio features into deep learning networks enabled the discernment of ND-infected poultry vocalizations. The development of such an automated detection system holds significant promise in enhancing animal welfare and the efficiency of poultry production monitoring.

\subsubsection{AI-assisted Food Synthetic Biology}
Different from traditional food production systems in agriculture and husbandry, synthetic biology can use cell factories or biosynthesis platforms to produce the food. It can use less land and water, does not use pesticides and fertilizers, and thus produces food in an efficient and environmentally friendly way~\cite{LV2021100025,jahn2023microbial}. For example, the brewing yeast cell factory is developed for efficiently synthesizing artemisinin, where the yield from the factory with 100 square meters is equivalent to the planting yield with nearly 50,000 mu~\cite{paddon2014semi}.

Massive data have been generated while studying the regulation principle of microbial cell metabolism and bioprocesses monitoring. In contrast with traditional methods, which adopt time-costing trial-and-error~\cite{gardner2013synthetic}, ML offers an opportunity to use these publicly available and experimental data to predict the impacts on the host and environment to greatly reduce unnecessary trials. AI has been applied to different aspects of synthetic biology, such as component engineering, genetic circuits and metabolic engineering~\cite{eslami2022artificial}. In component engineering, AI can improve the efficiency in functionally annotating biological components, accelerating the optimization speed of natural biological components, and enabling automatic design of genetic and protein components from scratch. For example, generative adversarial networks can be used to design functional protein sequences~\cite{repecka2021expanding} and promoters in Escherichia coli~\cite{wang2020synthetic}. In genetic circuits, AI techniques have been leveraged to predict the impact of constructs on the host and vice versa. For example, graph convolutional networks have been used to predict functions of proteins from protein-protein interaction networks~\cite{alipanahi2015predicting}. Sequence-based convolutional and recurrent neural network models have been used to design new biological constructs~\cite{eastman2018solving}. In metabolic engineering, AI has been applied to almost all stages of the bioengineering process, from annotating protein function, predicting synthetic pathways to optimizing the expression level of multiple heterologous genes~\cite{lawson2021machine}. AI and synthetic biology complement each other and will have great potential in high-quality, low-cost and large-scale synthesis of raw food materials~\cite{specht2018opportunities}.

\subsection{Food Processing and Packaging}
The food processing and packaging are mainly responsible for making foods accessible and desirable. For example, how to process food materials to generate delicious food with different flavors while retain the nutrient for consumers? How to improve value-added processing and efficiency in resource utilization to reduce food loss and waste? Traditional methods resort to a lot of human labor and do not fundamentally break free from the constraints of human experience, and thus are less efficient.

The key problems in food processing are the complexity and nonlinearity of the process, an extensive range of parameters, complex and various food compositions and structures, etc~\cite{datta2022computer}. AI provides one powerful tool for these problems for its great nonlinear fitting capability. According to the defined procedures of food processing, AI can improve production efficiency in primary and deep food processing, followed by food packaging via monitoring the food quality and safety, reducing food loss and waste, and improving value-added processing and resource utilization efficiency. Furthermore, with the development of AI, food processing and manufacturing are moving to intelligence and greenness, resulting in food intelligent manufacturing, which will also be discussed.

\subsubsection{AI for Primary Food Processing}
Primary food processing turns agricultural products, such as raw wheat or livestock, into edible ingredients or food products. In the sector of primary food processing, most food production fields have been affected by AI technologies such as meat, grains and fruit~\cite{mavani2021application}. For example, utilizing robotics with ML algorithms reduces the transmission of viruses in meat processing plants~\cite{hobbs2021covid}. For grains processing, ANN can prevent dust explosion by predicting the probability of explosion and monitoring the actual production environment~\cite{arshad2021modelling}. ML and CV methods have also been employed in meat quality assessment~\cite{penning2020machine}, authenticity detection~\cite{parastar2020integration} and fat content prediction~\cite{fowler2021partial} in meat and poultry processing industries. Through CV and infrared technology, multi-dimensional detection is carried out to classify the products with detected internal diseases and defect information for the fruit quality~\cite{wang2021advanced}. In addition, various primary food processing procedures have also been enhanced by AI, such as drying and sorting~\cite{mavani2021application}. For example, CV-based food sorting systems can automatically classify food according to the patterns of appearances, such as color, shape and texture to save time and money in the production line~\cite{dhiman2022fruit}. Combined with industrial large-scale datasets, DL methods have achieved promising performance in food image analysis, supporting further applications in food processing like quality monitoring and component assessment~\cite{Janmenjoy-IFP-CSR2020}.

\subsubsection{AI for Deep Food Processing}
Deep food processing refers to processing procedures that create ready-to-eat foods from pre-processed ingredients. Similar to primary food processing, AI has imposed a positive transition on various food industries of deep processing, such as the protein industry, sugar industry and ultra-processed food industry~\cite{Raj-CRFSN-2020,yang2021integrated}. Different from primary food processing, deep food processing involves more complex and non-linear processing and the change of physiochemical characteristics of food. The DL methods have the potential to model such nonlinear complex processing, such as baking, canning, soaking, filtration and encapsulation~\cite{Raj-CRFSN-2020,yang2021mechanistic}, which are hard to solve by traditional chemometric methods. As one example, speeding up infrared heating while keeping the nutritious value of food is challenging via traditional physical models. Artificial Neural Network (ANN) can predict the parameters of infrared heating to improve the efficiency~\cite{ogundele2022machine}. As an ultra-processed food with ingredient-intensive characteristics, the Bayesian model can better predict food properties to improve the efficiency of microwave-food product design, such as thickness optimization for better heating uniformity~\cite{yang2021integrated}. Moreover, ML methods can improve the process efficiency. For example, in traditional sugar manufacturing, samples are taken to the laboratory for offline analysis, leading to inefficient process monitoring due to information delay. In contrast, ML can deliver real-time information about the process streams and speed up manufacturing via analyzing online sample properties based on precise models~\cite{da2022handheld}. Additionally, various ML methods are developed for intelligent sensing of food flavor~\cite{zhang2019sour}, and are further used to regulate and develop engaging food flavors for consumers~\cite{ji2023recent}.

\subsubsection{AI for Food Packaging}
After food processing, the next step is food packaging, which includes the procedures for food protection. AI technologies empower intelligent food packaging both in terms of packaging procedures (e.g., filling and labeling) and packaging materials (e.g., organics from byproducts)~\cite{gonzalez2022novel}. In food packaging, CV-based object detection models can inspect and remove hazardous foreign objects for foreign object detection to increase the shelf-life of food~\cite{chen2001food}. To ensure the quality of food package sealing, a DL-based air-tightness diagnostics system is developed~\cite{polyakov2020industrial}. Traditionally, the date of food labeling is checked by manual operators, which is costly and prone to errors. DL approaches can realize the precise recognition of expiry date in retail food labeling to ensure the quality and safety of food in packaging~\cite{ribeiro2018end}. In addition, ML model is also used for selecting and designing the packaging materials as they can predict the best packaging properties. It can improve the efficiency in packaging material research to find more sustainable packaging materials~\cite{patnode2022synergistic}. For example, the KNN algorithm is applied to classify and select biodegradable active fish gelatin packaging derived from the by-products of fish processing~\cite{e2021artificial}. As resource-intensive manufacturing industries, food packaging produces daily waste, posing threats to the environment. As one solution to address the waste issue, ML algorithms empower food package recycling procedures (e.g., package sorting and automatic plant controlling) with the help of optical sensors data~\cite{kroell2022optical}.

\subsubsection{Food Intelligent Manufacturing}
Food intelligent manufacturing integrates the perception, control and decision-making enabled by AI to promote the intelligence of food design, production control, and manufacturing equipment to achieve precise, intelligent, and green food processing and manufacturing.

Food digital design focuses on modeling the correlation among food composition, structure, quality, and processing parameters based on the theories of physical, chemical, and biological changes in food processing. Most methods resort to physics-based or mechanistic models to simulate food transformations (e.g., from raw to dried) during processing. This type of method generally combines multiple physics and thus is hard to compute via conventional numerical simulation methods. In addition, various parameters associated in food processing should be known before, which are sometimes harder to meet in real scenarios. The new paradigm of AI for Science is emerging to solve these problems by providing large-scale data-driven modeling~\cite{zhang2023artificial}. Correspondingly, various optimization methods can be designed for effective model training~\cite{madoumier2019towards}. The future trend is to combine physics-based and data-driven models for food processing and manufacturing by combining their respective benefits, which is also another type of AI for Science~\cite{zhang2019hybrid}. In intelligent production control, AI can manage the process of food production to achieve high automation, such as automatic food quality monitoring and sorting~\cite{van2008kinetic,hassoun2023food}. Intelligent manufacturing equipment refers to the equipment with the capacity of perceiving, analyzing, reasoning and decision-making, such as food industry robots and food production lines. These types of equipment surpass manual methods in yield rates and reduce the physical stresses of human workers. For example, robotic systems are designed for various tasks like cutting, deboning, and packaging to improve efficiency and reduce labor costs in poultry processing facilities. Researchers~\cite{ahlin2022robotic} pioneered systems that capitalize on deep learning advancements in robotics. These systems use robot arms equipped with vacuum grippers and force sensors to securely hold chickens, paired with waterjet cutter arms to precisely trim and debone the chickens. However, modeling the complex dynamics involved in robotic collaboration and cutting of biological tissues remained an open challenge. To navigate this, a data-driven dynamics model~\cite{mitsioni2021modelling} can be used to capture the nuanced dynamics of poultry processing. This model employs LSTM networks and a model predictive controller, demonstrating strong potential for comprehending complex interactions compared with traditional models in robotic poultry processing.

As one trend in this field, digital twins can create a virtual mapping from physical spaces to digital factories to adjust food processing and optimize production parameters, opening new possibilities in upgrading the food processing and industry~\cite{VERBOVEN202079}. In addition, AI-enabled 3D food printing~\cite{escalante2021advances} can print foods with different attributes, such as shapes, texture and nutrient content, which has the potential to produce personalized foods to satisfy various personalized requirements.

\subsection{Food Transporting and Storing}
Food transporting and storing is the key intermediary between the rest steps involved in the food supply chain. Improving the efficiency of food transportation and monitoring to reduce food loss and waste is vital for satisfying food accessibility of consumers. In this section, we evaluate  AI solutions for food transporting and storing from five aspects, namely supplier selection and distribution planning, fleet scheduling and cooperation, traceability in food supply chains, demand prediction and dynamic transportation, and food preservation.

\subsubsection{AI for Supplier Selection and Distribution Planning} An appropriate setup plan for the food supply chain is fundamental to large food transporting systems involving multiple participants and large capital to reduce food loss and waste. The central issues of the plan are supplier selection and distribution planning. 
Food supplier selection involves comprehensive comparison between suppliers, and requires identifying key factors (e.g. product quality and transportation cost) and making feasible decisions under constraints~\cite{zhangModelCombiningBayesian2019}. In contrast with complex mathematical models like multi-criteria decision-making models~\cite{tirkolaeeIntegratedDecisionMakingApproach2021}, ML requires minimal expertise~\cite{zhangModelCombiningBayesian2019, stephenDecisionSupportModel2016}. For example, one can simply find potential influencing factors with a Bayesian network, and use the genetic algorithm to filter them~\cite{zhangModelCombiningBayesian2019}. 
As for distribution planning, conventional works like fuzzy models depend heavily on manually designed frameworks and often overlook implicit factors~\cite{peidroFuzzyOptimizationSupply2009}. However, various ML tools can mine implicit association rules to make a better
distribution plan~\cite{tingMiningLogisticsData2014, tufanoMachineLearningApproach2022}. For example, association rule mining can discover interesting patterns in wine delivery records, which helps determine the best delivery conditions and make an appropriate distribution plan~\cite{tingMiningLogisticsData2014}.

\subsubsection{AI for Fleet Scheduling and Cooperation} With a food supply chain built up, a fleet schedule requires to be outlined to improve the efficiency of transportation. Most conventional methods overlook time limits when transporting perishable food products~\cite{agustinaVehicleSchedulingRouting2014}, while ML methods consider time constraints by modeling the limited time window into fleet scheduling system, and using a genetic algorithm to accelerate the search for global optimal solutions~\cite{wangMultiobjectiveOptimizationDelivering2018}. 
As for cooperation in the food supply chain, there are also some attempts to improve known frameworks with different ML techniques (e.g., LSTM and reinforcement learning)~\cite{maoCreditEvaluationSystem2018, tangCooperationMultistageGame2005}. For example, LSTM can aggregate information from blockchain contracts to analyze events in the supply chain and give suggestions to regulators~\cite{maoCreditEvaluationSystem2018}. Overall, ML algorithms integrate time limits into the transportation system to make fleet scheduling more suitable for perishable food transportation with less manual inspection.

\subsubsection{AI for Traceability in Food Supply Chains} Food tracking and quality assurance are the top priorities in food product delivery to minimize the food loss and waste, where two aspects are taken into account, namely comprehensive transportation records and quality examination of food. ML methods can utilize food product information collected by sensors and generate more accurate transportation records based on the collected details~\cite{alfianImprovingEfficiencyRFIDbased2020, wongTechnicalSustainabilityCloudBased2021d}. For example, the ML method XGBoost is utilized to recognize passive RFID tag movements based on attributes extracted from sensors to decide whether the food products are shipped from or received by a warehouse, and automatically make real-time food transporting records~\cite{alfianImprovingEfficiencyRFIDbased2020}. ML is also combined with blockchain and knowledge graphs to uniform food product identifiers, maintain data integrity and preserve user privacy~\cite{wongTechnicalSustainabilityCloudBased2021d, pizzutiMESCOMEatSupply2017}. Meanwhile, quality examination of food requires in-time deterioration detection and food shelf-life prediction. Traditional methods continuously detect food deterioration via sensors and consume much electric power. In contrast, ML provides a more efficient and environment-friendly solution through deterioration mechanism modeling where main factors of food deterioration are modeled and analyzed~\cite{shahbaziProcedureTracingSupply2021}, or food quality prediction where food quality is tracked through the food supply chain and utilized for time series modeling~\cite{balamuruganIotBasedSupply2020}. AI-enabled traceability systems with big data and blockchain techniques can monitor the food safety by delivering vital information that is related to both food chains and foodborne virus events~\cite{zhao2021analysis}.

\subsubsection{AI for Demand Prediction and Dynamic Transportation} An intelligent food transportation system should be sensitive to customer demands and automatically adjust itself, thus preventing food loss and building a sustainable food supply chain. Because conventional works (e.g. risk management systems) require consistent monitoring of food delivery status, they lack preparation for potential delivery incidents and lag behind the consistently changing customer demands, resulting in food supply shortage or surplus~\cite{okeManagingDisruptionsSupply2009}. By contrast, ML has the ability to predict possible incidents and demand changes in advance~\cite{nagarajanIntegrationIoTBased2022, kappelmanOptimalControlDynamic2021}. For instance, big data mining and the Markov model can help the food system make a series of global optimal decisions, where the Markov model synthesizes history information and deduces future outcomes~\cite{kappelmanOptimalControlDynamic2021}. Meanwhile, other works pay attention to the prediction of incidents and frequent customer demand changes, where ML methods can foresee the disruptions during food transportation and avoid most of unnecessary loss~\cite{luangkesornAnalysisProductionSystems2016, mishraUseTwitterData2018}. For example, Bayesian model is utilized to analyze incidents in the food supply chain, and determine posterior distribution of incidents so regulators can run simulations and dynamically adjust the transportation before the incident happens~\cite{luangkesornAnalysisProductionSystems2016}.

\subsubsection{AI for Food Preservation} Appropriate storage environment and food quality surveillance play a significant role in prolonging the shelf life of food products, thus reducing food loss and ensuring food freshness during transportation. To maintain a better preservation environment, ML methods are widely utilized for food quality monitoring and food deterioration detection. For food products in stock, ML tries to determine their quality in a swift and non-destructive way, like  ANN for egg storage time determination via a near-infrared (NIR) reflectance spectrometer, and SVR for avocado ripeness judgment with RGB images~\cite{coronel-reyesDeterminationEggStorage2018, choDeterminationHassAvocado2020}. With ML models built across various food preservation conditions, researchers can even gain more insight into best storing conditions (e.g., temperature and duration of storage), which can be utilized to adjust preservation environment~\cite{gorzelanyModellingMechanicalProperties2022}. Meanwhile, ML can also detect possible deterioration of stored food and alert people to the disruption to avoid food loss, such as analyzing RGB and NIR data with KNN to detect potato sprouting, and using partial least squares regression, PCA and principal component regression to detect possible food contamination~\cite{radyComparisonDifferentOptical2020, soniHyperspectralImagingMachine2022}.

\subsection{Food Marketing and Retailing}
Food marketing and retailing contain activities that link the food producers with food consumers, and they are responsible for the accessibility and desirability of food products for consumers~\cite{FoodMarketing}.
Through food marketing, food products and corresponding services are promoted among targeted customers, whose feedback is collected and reviewed in turn to fulfill more potential consumers' demands.
Through food retailing, food products are managed to ensure freshness and provided for consumers at appropriate price. In this section, we introduce how AI can enable food marketing and retailing to provide sufficient, nutrition-rich food products, and empower consumers to make more informed food choices, fueling rising demand for sustainable healthy diets.

Different from traditional approaches, AI can analyze the big data from customers’ purchasing data to predict the expected outcomes of improved products and provide valuable suggestions for producers to satisfy the customers’ demands. As one example, marketers can first utilize clustering approaches to group consumers by similarity and determine targeted customers before transactions~\cite{dekimpe2020retailing}. Then based on sales data, marketers can collect target customers' feedback from the social network with social listening technology for further analysis~\cite{brooks2022use} and also determine crucial product attributes (e.g., whiteness and smoothness of yogurt) with autoencoder-based optimizing approaches~\cite{bi2022consumer}.
Finally, they can predict the expected outcomes of improved products with supervised learning approaches and provide valuable suggestions for producers~\cite{dekimpe2020retailing}. In addition, big data from the Internet and social media can be mined by AI to better regulate advertising and marketing to positively influence individual food choices for their health. 

Retailers' responsibilities include ensuring product access to the consumers, which are essential for heightened consumer satisfaction~\cite{verma2021artificial}. 
AI can provide convenience for retailers, such as dynamic~pricing~\cite{burman2021deep}, sales forecasting~\cite{tsoumakas2019survey}, and inventory management~\cite{hou2021foodlogodet} to satisfy user's demand while reducing the food waste. The dynamic pricing policies for food products are related to many factors, such as freshness, sales, stocks, and competitor strategies~\cite{verma2021artificial}. Among these factors, freshness~\cite{guo2020portable} is the top priority, especially for perishable food, which is unfortunately overlooked by traditional supply chain models~\cite{huang2021optimal}. In contrast, AI can effectively utilize the food expiration date as a clue for dynamic pricing~\cite{yang2022dynamic}. For example, one deep reinforcement learning method deep Q-Network is introduced to predict customer demand function with the shelf life and price as the input, and it is further utilized to determine the optimal pricing strategy~\cite{burman2021deep}.
In addition to dynamic pricing, sales forecasting can also help retailers get a hold of consumers' demands and reduce stocked products. 
In the past, the solution to this problem often heavily relied on human expertise.
Even though some conventional software can help, their models only simply make predictions based on the average of past sales data.
ML methods can take both internal data (e.g., past sales data and product information) and external data (e.g., weather) into account to reduce the error from unknown causes and can further improve the accuracy with the increase of the amount of input data~\cite{tsoumakas2019survey}. Regarding inventory management in retailing stage, recognition methods such as logo detection~\cite{hou2021foodlogodet} and QR code detection~\cite{bhatia2023blockchain} can efficiently help check and manage food products on shelves.

All these above-mentioned activities involve food disposing, which is the management of food loss and waste (FLW) along the food supply chain, such as agricultural residues, processed scraps, degraded food and kitchen waste. Therefore, we also introduce how AI can improve food disposing in the next section.

\subsection{Food Disposing}
AI can promote the management of FLW at different stages to enable more sustainable healthy diets by improving the availability and accessibility of food. From predictive algorithms that harmonize food production with actual demand to advanced techniques that ensure quality control in food processing and transportation, AI plays an instrumental role in reducing FLW. At the consumption end, the integration of AI tools can transform household storage and kitchen waste management to ensure efficient resource utilization and minimal wastage.

\subsubsection{AI for FLW in Consuming} 
Food waste has become a serious problem in the consuming stage where customers finish purchasing, preparing and eating in households and restaurants. 
Household food storage is blamed for the majority of waste due to the lack of monitoring on the fresh state of the food. Detection models can tell consumers if the food is fresh for accurate food monitoring to reduce waste~\cite{Sanghi2022fresh}.
Another key issue is the management of kitchen waste. A complete kitchen waste processing procedure involves collecting, sorting and decomposing procedures. In the collecting process, AI models for accurate prediction of waste generation and diversion based on socio-economic variables are developed for intelligent waste management~\cite{Kan2017MSW}. 
Sorting waste after collecting is vital for resources in kitchen waste. Waste recognition algorithms are proposed to automatically sort food waste in trash bin~\cite{Zhang2021MSW}. 
When sorting is finished, composting kitchen waste is the best way to recycle resources. 
By applying ML-based models, it is feasible to measure and evaluate complex decomposing processes. As one example, CNNs are introduced for fast evaluation of compost maturity by analyzing images of composting stages~\cite{seo2021prediction}.

\subsubsection{AI for FLW in Production, Processing, Transportation and Marketing} 
In food production, excessive products from overproduction easily lead to FLW. 
The key role of AI is to help plan production and harvest to ensure stability between demand and actual output. For example, the RNN-LSTM algorithm is proposed to fit historical yield data and predict deviations in production based on predicted yields to reduce food loss from overproduction~\cite{Reb2019Banana}. 
Food losses during processing primarily result from substandard ingredients, poor quality control and unsustainable processing techniques.
As an example, to deal with substandard ingredients, CNN-based fruit defect detection and grading systems are used to reduce food loss by automating production lines~\cite{tbk2022detect}. 
For handling spoilt food from poor quality control, various vision techniques for finding spoilt are proposed. For example, hybrid algorithms are used to provide objective color recognition for spoilt food monitoring~\cite{lavika2020compute}.
Sustainable processing techniques on AI could significantly reduce food loss. For example, 
one ML framework is applied to turn food waste into a useful `by-product', where PCA is used for selecting of vegetable sources with potential prebiotic activity to produce carbohydrate structures and fibres~\cite{sab2021by}. 
A significant factor leading to transportation loss is improper cold-chain conditions during transport.
To address this issue, various AI tools are used in cold-chain condition monitoring. For example, food recognition algorithms are used to monitor food in the refrigerator and support appliance management, which ultimately reduces energy consumption and potential food loss~\cite{bonaccorsi2017highchest}. 
In food marketing, waste can be caused by inappropriate prices that bring the demand contraction and the expiry of unsold products. By offering dynamic pricing based on remaining shelf life, AI models can reduce waste and effectively connect price-demanding consumers with products, thereby reducing food waste~\cite{burman2021deep}.

\subsection{Discussion}
AI enables every stage of food systems to make sustainable and healthy diets available, accessible, affordable and desirable to all consumers. For example, AI can foster sustainable productivity growth and diversity via  AI-enabled precision agriculture to promote food production in sufficient quantities and diverse types. AI can effectively reduce FLW at different stages of food systems to support the accessibility of food. AI can help better regulate advertising and marketing to help consumers make more informed dietary choices. 

Of note, fully harnessing the potential of AI technologies for sustainable food systems requires the right enabling environment, such as favorable policies and regulations, adequate energy, public research and development, taxation and subsidies, especially in the stage of consumption. Consumption involves more social, political, cultural and individual factors, and it is not simply enabled by AI technologies. For example, there are many factors to influence the diet choice of people in food consumption, such as their income and product price, taste, nutrition, and social and cultural norms, which probably do not make people select sustainable healthy diets. When AI uses these purchasing data to promote products and provide recommendations, and the recommended results in turn enhance the negative loop. In this case, policymakers should focus on regulations to restrict companies to encourage people to make better dietary choices. Another example is that healthy diets are unaffordable for over three billion people due to their bad economic conditions and geographic constraints. The fast development of AI  often appears in developed countries and regions. From the global aspect, the uneven use of AI further stops current food systems from moving to globally sustainable ones. How to make AI technologies to increase the productivity of food and reduce food loss and waste in these poor regions is a formidable challenge.

On the other hand, the effective implementation of these right enabling environments can also benefit from the utilization of AI. Take the policy formulation as one example, AI can analyze the food quality to make public health policy~\cite{bourguet2013artificial}, and predict consumption trends to make policies that direct the amount of identified possible production to minimize food waste~\cite{fabi2021improving}. The ML approach can predict the food security status from secondary data, such as the prevalence of people with insufficient food consumption. The estimated results can be used for governments to make timely adjustments to relevant policies~\cite{martini2022machine}. Another example is the energy in the food systems. AI could provide technical support for developing renewable energy or food-related energy-efficient appliances. For example, CNNs could be used to monitor and optimize the digesting process in biogas production~\cite{seo2021prediction}. Recognition algorithms can be embedded in refrigerators to identify food and support cold-chain management, which ultimately reduces energy consumption from overcooling~\cite{bonaccorsi2017highchest}.

In addition, food systems are the largest employment sector, especially in high-income countries. The integration of AI into the food systems leads to reduced human labor. Some mechanical and repetitive tasks are gradually being replaced by automation in the agriculture and food industries. The changes in the labor force structure have brought new employment challenges to the AIFS. Meanwhile, AIFS can also bring new employment opportunities, such as agricultural consultants and professional service providers, who are highly skilled in plant production and protection, robotics, farm management and business to ensure foods move along value chains more efficiently and improve the food accessibility~\cite{munnisunker2022impact}. With proper adjustment, a reasonable redistribution of human labor could be achieved to promote efficient human-machine cooperation in AIFS.

\section{Challenges and Future Directions}
\begin{figure*}[htbp]
\centering
\includegraphics[width=1\textwidth]{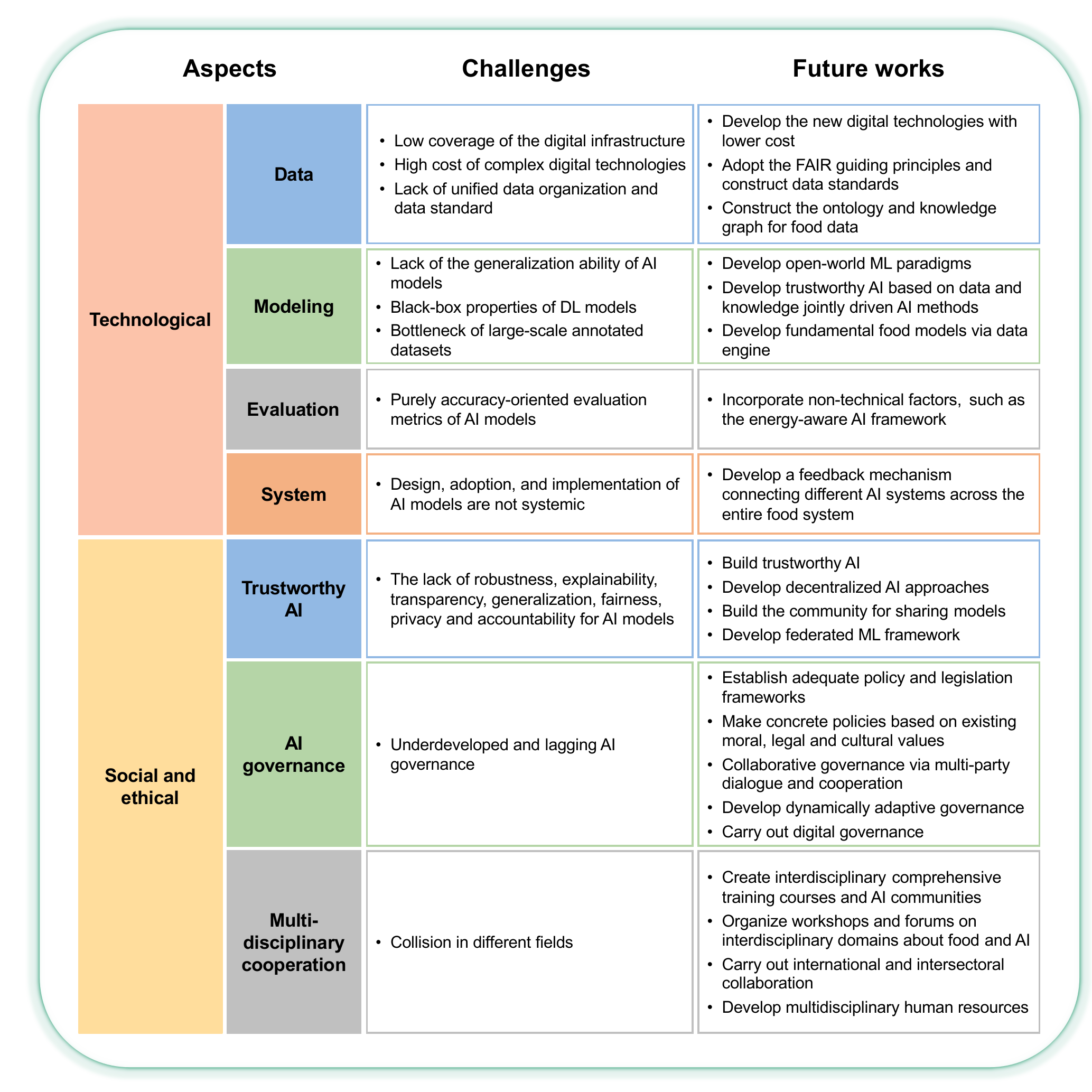}
\caption{Challenges and future directions of AIFS.}
\label{ChallFutDir}
\end{figure*}

Despite of great potential of AI in accelerating positive transitions of consumer-driven food systems to make the food universally available, accessible, affordable, and desirable for humans toward sustainable healthy diets, current applications of AI technologies in food systems are still lagging behind. This section identifies key challenges and future works in AIFS from technological, social, and ethical dimensions, which are illustrated in Fig.~\ref{ChallFutDir}.

\subsection{Technological Dimension}
The technical challenges of applying AI to food systems include the low use efficiency of digital technology, the lack of unified data organization and standards, the non-robustness of AI models, and insufficient metrics for comprehensive model evaluation. These are mainly illustrated by data, modeling and evaluation aspects. Finally, we discuss the technological dimensions of the whole food systems, since our final goal is to increase the efficiency of entire food systems.

\textbf{Data Aspect.} An important driver for applying AI to food systems is the large amount of data generated from the digitalization of food systems. Recently, there has been a consistent emergence of extensive food datasets built for food recognition and plant disease detection~\cite{min2023large, liu2021plant}. However, many stages of food systems are underserved by the computing infrastructure. For example, the digitalization of farms and supply chains is still in its infancy especially in low- and middle-income countries (LMICs), where there is no internet access or IoT coverage in many areas~\cite{mehrabi2021global}. Complex digital technologies are not yet widely used due to costly equipment and high maintenance costs. Even though big data from food systems are available, it is hard to directly utilize these data without the unified data organization and standard due to their different structures~(e.g., unstructured and semi-structured ones), different modalities~(e.g., different types of imagery) and different sources~(e.g., drones and satellites) and different scales~(e.g., food composition, food structures and food appearance information )~\cite{costa2019opportunities}. In addition, much data with low quality is incomplete, messy, or biased, making ML methods provide predictions with lower accuracy and sub-optimal decision-making results.

New low-cost digital technologies should be developed to be affordable for all the actors in the food systems. Based on the data generated from these digital technologies, data standards construction is vital to handle data challenges. The FAIR guiding principles should also be adopted in the food systems to fully maximize the value of data~\cite{wilkinson2016fair}. With respect to the collection and integration of data, the shared mechanism and unified data interface standards can be negotiated between government and business. Moreover, LLMs are expected to lift the burden of unifying data standards in the future, as different formats of data can be transformed into a unified form processed by Large Language Models (LLMs)~\cite{zhao2023survey}. A food-specified GPT model should be developed to handle the scenario-specific data and format generated in the food systems. Additionally, Artificial General Intelligence (AGI) based on LLMs is able to interpret the complex raw data into analysis results using understandable language to actors in food systems~\cite{denkenberger2022long}, e.g., farmers~\cite{lu2023agi} and corporate decision makers\footnote{https://tastewise.io/blog/introducing-TasteGPT}. Furthermore, the food ontology and knowledge graph can provide a standardized conceptual terminology to connect the data in the whole food systems in one structured way, and thus can effectively organize these huge volumes of multidisciplinary and heterogeneous food data to ease the problems of data harmonization and interoperability and support the utilization of AI technologies across disciplines and sources that differ in modality, structure and scale~\cite{dooley2018foodon, MinPattern2022}. Setting a standardized, safe and controllable national and international public data open platform is necessary to promote the development and utilization of public data resources in an orderly manner. 

\textbf{Model Aspect.} A major challenge faced by most AI models stems from their lack of generalization ability to different environments. The conventional ML paradigm generally assumes close-environment scenarios with important factors of the learning process invariant. However, food systems often involve an open and dynamic environment,  such as the farming environment with various production conditions. What's worse, data are usually accumulated with time, like endless streams in real-world scenarios, making existing ML models almost impossible to train after collecting all the data~\cite{Zhou-nsr-nwac123}. Therefore, conventional ML algorithms are often not practical when they are applied to food systems. Moreover, food systems are non-linear and complex, especially in food processing and manufacturing, the processing involves complex correlations among food processing conditions, changes in component structure, and food quality and functional characteristics. How to model such complex interactions and predict physiochemical properties is very challenging. In addition, DL has been widely applied in food systems~\cite{zhou2019application}. However, current DL models are mostly considered black-box models without a clear understanding of how the algorithm generates predicted recommendations. They also require large volumes of labeled, curated, high-quality data for training, which has become a bottleneck in many application scenarios. For example, the large-scale annotated dataset may not be available for new fish species in aquaculture~\cite{yang2021deep}. 
Furthermore, future food systems are consumer-centric and thus how to utilize various characteristics of person, from their genes, gut microbiome, metabolism, food environment to food preference and nutrition to build the personal food model is also important~\cite{10.1145/3394171.3414691}.

In this regard, AI technological innovation will play an important role from the following several aspects. 

\begin{myitemize}
 \item The first priority is to develop open-world ML methods, and some open-world ML techniques such as incremental learning and continual learning on emerging novel classes have been explored~\cite{Zhou-nsr-nwac123, gu2021open}. There have been some research works on how to make the model with the ability of lifelong learning via online continuous learning to cope with the stream data. 
 \item The food processing is one complex and ever-changing nonlinear reaction process, and the food structure also exhibits complex dynamics. Single-dimensional perception technology can only obtain limited information during the process, and cannot accurately control the food processing process. In order to obtain more comprehensive and reliable information, it is necessary to integrate multi-sensory information to compensate for the shortcomings of single perception technology  via various sensing technology and the expanded sensing range. Food forms a complex system with multi-scale structures, and the processing process is influenced by numerous factors. Food digitization and AI provide effective tools for exploring the underlying interaction mechanisms and patterns, helping to accurately model the impact and correlation of food processing on multi-scale structures (such as molecular chain structure and aggregation behavior) and interactions (such as inter-molecular association and aggregation). 
 \item Existing purely data-driven AI methods are neither explainable nor robust. It is essential to develop data and knowledge jointly driven AI technologies, which embed human knowledge in the development of AI models to cope with situations that were not foreseen at the training stage. For example, the Chain of Thoughts (CoT) and Reinforcement Learning from Human Feedback (RLHF) techniques applied in ChatGPT\footnote{https://openai.com/blog/chatgpt/} bring a new perspective on integrating human knowledge into LLMs. Furthermore, we should develop trustworthy AI to make the AI models with better robustness, explainability and transparency~\cite{li2021trustworthy}, which will be detailed in the section on social and ethical aspects. In addition, fundamental models by large-scale pretraining like the Segment Anything Model (SAM) are proven to have promising generalization ability and zero-shot ability. However, their performance under specific vertical scenes is challenged~\cite{zhang2023comprehensive}, e.g., food scenario. Food-specific LLMs have enormous potential in handling real-world issues under complex open-world environments. The concept of data engine applied in those methods can alleviate the reliance on large-scale annotated datasets. With the developed fundamental models with reliable generalization ability, effective data annotators can be developed based on these models for large-scale annotations, thus forming a virtuous cycle of data.
 \item In consumer-driven food systems, building personal food model is necessary. Wearable and mobile sensors can collect various personal data with different aspects, such as nutrition and metabolism, resulting in personal big data. AI, especially ML provides one powerful tool for integrating these heterogeneous data for personal food modeling. The personal food model can be further used for analyzing personal dietary and flavor requirement, and predicting their health status.
 \item At last but not least, we should bridge gaps between technology innovation and user adoption. For example, simple-to-use technologies that increase farmers’ yields and profits are often adopted rapidly. More complex approaches that may have longer-term benefits but do not necessarily increase farm profits immediately are often adopted much less rapidly without specific extension and training efforts~\cite{FFS-Report2020}.
\end{myitemize}

\textbf{Evaluation Aspect.} AI should be approached from a socio-technical perspective that simultaneously considers environmental and economic factors and other non-technical factors~\cite{nishant2020artificial}. However, existing performance-driven evaluation metrics of AI models are insufficient for food systems without considering other factors, such as the impact of models on environmental sustainability. Therefore, appropriate metrics must go beyond purely accuracy-oriented technical dimensions and should be re-designed to consider more factors, such as energy consumption, resource utilization and various dimensions of AI trustworthiness. For example, the energy-aware AI framework can be developed to jointly optimize the energy cost and the performance of AI models~\cite{chen2022survey}. 

\textbf{System Aspect.}
In our innovative concept, the function of AI cannot be viewed in isolation. It is intrinsically embedded within the broader ecosystem of AIFS. The design, adoption, and implementation of AI models in food systems are often confronted with issues that are not just technical but systemic. For instance, the integration of various AI components within a given system may introduce unforeseen complications (from redundant duplicate modules) and inadaptability (from mutually exclusive interfaces). The AI component that may work well in one stage of the food supply chain, says precision agriculture, may not transfer well into another, such as food retailing management. Moreover, the dynamic nature of food systems, influenced by variables like climate patterns, socio-economic conditions, and geopolitical scenarios, demand AI solutions that can adapt in real-time. Current systems, not designed with such adaptability in mind, can hinder the seamless integration of newer AI solutions. This calls for periodic system overhauls and upgrades, which can be both time-consuming and expensive. Lastly, the scalability and customization of AI solutions are crucial. While AI-driven solutions might be scalable, they should also be flexible enough to cater to local needs and conditions. A one-size-fits-all approach can lead to inefficiencies and even failures in addressing the unique challenges posed by different sections of the food system.

Therefore, the interoperability of AI solutions, especially when they span different domains or stakeholders within the food system, is paramount. 
There is also the need for a feedback mechanism wherein the output of one AI system serves as an input or modifier to another. This interconnectedness ensures that improvements or detriments in one section are promptly addressed in others, leading to harmonious and efficient AIFS.

\subsection{Social and ethical dimension}
Social and ethical challenges come from the need to build trustworthy AI, improve AI governance, strengthen transdisciplinary cooperation and promote intersectoral collaboration.

\textbf{Trustworthy AI.} There are many aspects of AI trustworthiness, such as robustness, fairness, privacy and accountability of AI systems~\cite{li2021trustworthy}. We will detail some aspects to show the challenges of AIFS. (1) Due to unclear and inadequate rules on data governance, many fundamental issues on data privacy, such as data ownership and utilization have to be addressed~\cite{panch2019inconvenient}. The lack of decentralized data management and  ML systems further decreases the willingness of multiple actors or stakeholders to share data. Insecure data sharing between multiple groups may also impede fully exploiting the data and developing AI technologies. These issues lead to a number of barriers to the wide adoption of AI in food systems. (2) Current development of AI technologies is unevenly distributed around the world, and heavily skews toward more developed areas, increasing inequalities both between and within countries~\cite{zhang2021ai}. Even worse, some countries and intergovernmental organizations restrict AI technology exchanges to maintain their own AI technology advantages, leading to inequitable societies. (3) The sustainability for societal and environmental well-being should be necessary for AIFS. However, existing AI technologies such as DL require massive computational resources with a high energy requirement and carbon footprint~\cite{jones2018stop}. One example is recently developed LLMs for their amazing performance. However, larger models based on Transformer require more expensive computational power and higher consumption of training time, leading to a drastic increase of carbon footprint. For example, the carbon footprint of training a Transformer model is as high as 626,155 pounds of CO$_{2}$ emission~\cite{strubell2020energy}.

To solve these challenges, it is imperative to develop the ethical standards and legal framework regarding different dimensions of trustworthy AI~\cite{floridi2019establishing}. A number of influential organizations and research institutes, such as the International Organization for Standardization (ISO) and the European Union (EU) have been taking action to address the issue of trustworthy AI. Meanwhile, building trustworthy  AI models has been becoming an active topic of AI research in the AI community. Decentralized AI approaches are developing to make the implementation of AI technologies be adapted to the cultural background and particular needs of different regions for promoting societal equality~\cite{montes2019distributed}. Developing computation-efficient AI, such as pruning, quantization, and knowledge distillation has been developed to reduce the environmental impact of AI for sustainable AI~\cite{chen2022survey}. For the fairness of AI, recent works have proposed different methods, such as identifying sources of inequity and de-bias training data to reduce the bias~\cite{courtland2018bias}. Several types of AI models, such as model-based and example-based methods are developed for explainable AI to explain their capabilities and understandings~\cite{chen2022survey}. With the emphasis on data privacy and security, a federated ML framework can train  ML models in a  distributed learning scheme to enable participants to jointly train one ML model without sharing their private data~\cite{yang2019federated}. It is also essential to promote the development of novel methodologies to assess the societal, ethical, legal and environmental implications of new AI technologies prior to launching large-scale AI deployments in the food systems. 

\textbf{AI Governance.} While AI technology can yield positive impacts and benefits for food systems, it can also generate unexpected and unintended consequences, and pose new forms of risks that need to be effectively managed by governments. For example, as AI systems learn from data, unanticipated situations that the system has not been trained to handle can lead to unexpected dangerous behaviors. AI governance involves decisions about how AI should be adopted or regulated. Although many high-level ethics guidelines for AI have been proposed, concrete policies within the context of existing moral, legal and cultural values have not been made. Another point is that the development of AI strategies is unevenly distributed around the world, where most of the existing strategies have been launched in developed countries, presenting limitations with AI policy and governance development~\cite{ulnicane2022governance}.

The first step is to establish one adequate policy and legislation framework to help direct the vast potential of AI toward the highest benefit for the food systems. AI technological advances must be coupled with socio-cultural changes and policy support~\cite{barrett2020bundling}. Policymakers should have a sufficient understanding of AI challenges to formulate sensible policy~\cite{vinuesa2020role}. It is essential to ensure the usability and practicality of AI technologies for governments. This would also be a prerequisite for understanding the long-term impacts of AI, while regulating its use to reduce the possible bias inherent to AI~\cite{courtland2018bias}. Second, collaborative governance is necessary and requires multi-party dialogue and cooperation. In addition to the government, stakeholders from a wide range of public, private, and civil society sector actors need to be included in the governance process. The government should fully combine the advantages of all these participants. Third, the governance methods should be dynamic, changing adaptively with AI development. In the real-world situation, any progress in AI technologies can present huge challenges to current governance systems. Therefore, AI governance should maintain iterative optimization.  Rapid changes also propose higher requirements for the high-speed reaction ability of governance participants. Once new AI technologies are in place, supporting policies should be made, such as lower social costs to ensure the effective deployment of these AI technologies. In addition, existing data regulatory requirements and privacy concerns can impede the development of AI models for food systems~\cite{liang2022advances}. Therefore, digital governance is important. It is necessary to establish a system of data property rights, promote the classification and grading of public data, enterprise data and personal data to confirm and authorize the use of data, followed by establishing a property rights operation mechanism that separates the right to hold data resources, use data processing and manage data products. 

Besides these above-mentioned actions, the actors' acceptance of these emerging AI technologies should also be carefully considered. More science communication efforts should be made to help the public get a better understanding of AI technologies to address public concerns and prejudices against AI technologies~\cite{nsrnwac068}. 

\textbf{Transdisciplinary Cooperation and Intersectoral Collaboration.} AIFS has significant interdisciplinary characteristics, and sits at the intersection of AI, agriculture, food science and engineering, and other disciplines, leading to many challenges. One challenge comes from the need to blend expertise from different disciplines. This is because these scientists from different disciplines are trained differently. For example, the discipline of food science is typically far from providing computational engineering training and food researchers may have slower perception on AI progress. Those with advanced engineering training pay less attention to food applications due to fewer opportunities. AI researchers may suffer from the prior bias that AGI on general objects is primarily to be developed and neglect the significance of AI-enabled food research. Another example is the gap between farmers and AI researchers. Seamless translation from agricultural problems to decision-making models is crucial for successfully adopting AI in agricultural production. However, AI researchers usually do not know much about agricultural knowledge and are thus unaware of these AI-solvable agricultural problems. In addition, AI technologies for food systems may have important legal and ethical questions since they embody significant amounts of automation and intelligence. These questions span law, public policy and ethics, leading to the collision from different fields~\cite{Russell_Dewey_Tegmark_2015}. 

Therefore, AIFS requires transdisciplinary cooperation and intersectoral collaboration. First, AIFS requires the participants in the food systems to receive comprehensive training in both computational skills and other fields. A possible way is to construct a trustworthy AI community in the food industry~\cite{datta2022computer}, which involves creating computational resources like hardware, databases, and software platforms to share food-centered AI models and expertise. Based on the foundational food model repositories, cross-training courses can be further created to connect academics with industries to meet the evolving demands of the AIFS. For example, bench scientists in food industries are trained in ML while computational scientists in AI academia need to be trained in food science. Second, several community-wide efforts to support interdisciplinary research in AIFS are recommended, such as recently organized conferences in interdisciplinary domains, including the workshops of vision for agriculture 
associated with CVPR2022~\cite{CVPR2022},large-scale fine-grained food analysis associated with ICCV2021~\cite{ICCV2021}, AIxFood with IJCAI 2020~\cite{IJCAI2020}, and Multimedia Assisted Dietary Management with ACM MM2023~\cite{MM2023}, tutorial of food computing with ACM MM2020~\cite{jiang2020food}
Some advanced interdisciplinary institutes, such as the AI Institutes for Next Generation Food Systems are built to utilize AI to secure enough healthy, nutritious, and sustainable food for people and the planet. Third, scaling emerging AI technologies could have a major impact on food systems and requires an intersectoral collaboration ecosystem where governments, food companies, investors, donors and leaders work together to provide support to technology innovations across their life cycles. Additionally, the development of AIFS is not determined by any single country. Therefore, all the countries in the world should work together to build better technological solutions, and better serve the world's population by our shared humanity via intersectoral collaboration. 

\section{Conclusion}
We are at a critical turning point for the third wave of AI to enable consumer-driven food systems towards sustainable healthy diets. In this work, we discuss how AI is leveraged across the entire food system from fork to farm to improve the availability, accessibility, affordability and desirability of the food for consumers. Despite of great potential of AI technologies for food systems, they are not yet fully developed due to technological, sociological and ethical challenges. These challenges can be solved by constructing data standards, accelerating AI technological innovation, building trustworthy AI, improving AI governance and enhancing multidisciplinary cooperation. There is a great opportunity for academia, industries, government and international organizations to work together to improve the global food systems via AI. Additionally, AIFS does not merely involve technical issues. Using AI to accelerate the food systems transition toward desired states should need the synergies between AI technologies and non-technical factors, such as societal dialogue, governance, regulators and ethical frameworks.

\ifCLASSOPTIONcaptionsoff
  \newpage
\fi



\bibliographystyle{IEEEtran}
\bibliography{IEEE_barejrnl}
\end{document}